\documentclass{PoS}

\title{Exotic static 3-body potentials}

\ShortTitle{Exotic static 3-body potentials}

\author{
\speaker{Pedro Bicudo} \\
CFTP, Instituto Superior T\'ecnico\\
\email{bicudo@ist.utl.pt}
}

\author{Marco Cardoso \\
CFTP, Instituto Superior T\'ecnico\\
\email{mjdcc@cftp.ist.utl.pt}
}

\author{
Orlando Oliveira \\
CFC, Universidade de Coimbra\\
\email{orlando@teor.fis.uc.pt}
}

\abstract{
We study exotic static 3-body potentials, utilizing generalized 
Wilson Loops in SU(3) lattice QCD.
For the quark-antiquark-gluon techniques  we address the 
angles of 0, 45, 60, 90, 120, 135 and 180 degrees, between the quark-gluon and the antiquark-gluon
segments. We calculate the form of the static potential and discuss whether, or not, two-body interactions exist between the three different bodies, and study the existence of repulsion between the
strings. We also perform a first study of the interactions in the system of  three gluon.
}

\FullConference{
XXVI International Symposium on Lattice Field Theory, July 14 --19, 2008, Williamsburg, VA, USA
}

\begin{document}

\section{Motivation}

We explore, in Lattice QCD, the static potential of three-body systems with gluon(s),  using Wilson loops .  We study the quark-antiquark-gluon hybrid and the glueball system composed of three gluons. 
The interest in three-body quark-antiquark-gluon and gluon-gluon-gluon systems 
is increasing 
in anticipation to the future experiments BESIII at IHEP in Beijin, GLUEX at 
JLab and PANDA at GSI in Darmstadt, dedicated to study the mass range of 
charmonium, with a focus in its plausible excitations and in hybrid and
glueball production.  The three-gluon glueballs are also relevant to to the odderon.

Thus several models of hybrids and of three-gluon models have already started to be developed. 
While there are several evidences, both in lattice QCD, and with Schwinger-Dyson
or many-body techniques for a massive gluon, even for massless gluons, the knowledge of a static potential would at least provide one of the components
of the dynamical potential.

The Wilson loop method was devised to extract, from pure-gauge QCD, the static potential 
for constituent quarks and to provide a detailed information on the confinement 
in QCD. In what concerns gluon interactions, the first Lattice studies were
performed by Michael 
\cite{Michael:1985ne,Campbell:1985kp}
and Bali extended them to other SU(3) representations 
\cite{Bali:2000un}.
Recently Okiharu and colleagues 
\cite{Okiharu:2004ve,Okiharu:2004wy}
studied for the first time another class of exotic hadrons, extending the Wilson loop of three-quark baryons
to tetraquarks and to pentaquarks. Very recently, Bicudo, Cardoso and Oliveira  continued
the Lattice QCD mapping of the static potentials for exotic hadrons, with the
studies of the hybrid quark-antiquark-gluon static potential and, of the three-gluon glueballs 
\cite{Bicudo:2007xp,Cardoso:2007dc,Cardoso:2008sb}.

\section{The Wilson loops for exotic systems}

To measure the static potential in lattice QCD, we employ the well know technique of the Wilson loop.
In this technique, an operator that describes the state being studied is constructed.

In the case of mesons and baryons, we only had quarks and antiquarks, being their time propagation
represented by fundamental paths (products of fundamental links). Now, we also have adjoint
paths, corresponding to the gluon propagation, and are given by
\begin{equation}
	\tilde{U}^{ab} =  \frac{1}{2} Tr[ \lambda^a U \lambda^b U^{\dagger} ]
\end{equation}
where $U$ is the corresponding fundamental path.

\begin{figure}
\centering
\begin{picture}(350,100)(0,0)
\includegraphics[width=0.8\textwidth]{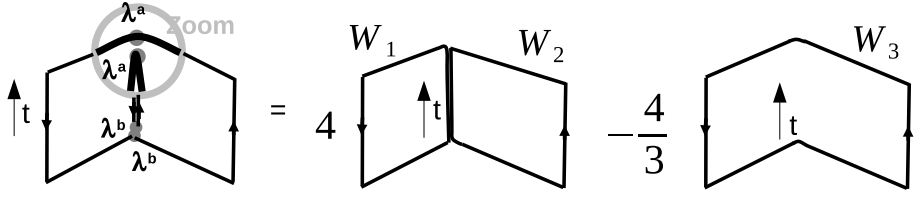}
\end{picture}
\caption{
Wilson loop for the hybrid meson system.
}
\label{loopqqg}
\end{figure}

The expressions can then be simplified by using the Fierz relation
\begin{equation}
	\sum_a \lambda^a_{ij} \lambda^a_{kl} = 2 \delta_{il} \delta_{jk} - \frac{2}{3} \delta_{ij} \delta_{kl}
	\label{fierz}
\end{equation}
For the hybrid system, the wilson loop is given by \cite{Bicudo:2007xp} (Fig. \ref{loopqqg})
\begin{eqnarray}
	W_{q\bar{q}g}= &&
	%
	\mbox{Tr} \Big\{ 
  	U^\dagger_4 (t-1,x) \cdots U^\dagger_4 (0,x)  ~  \lambda^b 
	\nonumber \\
  	&&
   	U_4 (0,x) \cdots U_4 (t-1,x)  ~ \lambda^a \Big\} ~ \times
	\nonumber \\
	\mbox{Tr} 
	\Big\{ &  &
   	U_{\mu_2} (t,x) \cdots U_{\mu_2} (t,x+(r_2-1)\hat{\mu}_2) 
	\nonumber \\
  	&&
   	U^\dagger_4 (t-1,x + r_2 \hat{\mu}_2) \cdots 
                 U^\dagger_4 (0,x+r_2\hat{\mu}_2) 
	\nonumber \\
  	&&
   	U^\dagger_{\mu_2} (0,x + (r_2-1) \hat{\mu}_2) \cdots 
        U^\dagger_{\mu_2} (0,x) ~  \lambda^b 
	\nonumber \\
  	& &
   	U^\dagger_{\mu_1} (0,x - \hat{\mu}_1) \cdots 
                 U^\dagger_{\mu_1} (0,x - r_1 \hat{\mu}_1 )
	\nonumber \\
  	&&
   	U_4 (0,x - r_1 \hat{\mu}_1) \cdots 
                 U_4 (t-1,x-r_1\hat{\mu}_1)
	\nonumber \\
  	&&
   	U_{\mu_1} (t,x - r_1 \hat{\mu}_1) \cdots 
                 U_{\mu_1} (t,x - \hat{\mu}_1 ) ~ \lambda^a \Big\} ~ .
\end{eqnarray}
where $r_1\hat{\mu}_1$ and $r_2 \hat{\mu}_2$ are the positions of the quark and the antiquark in relation to the gluon.

Using (\ref{fierz}), this operator becomes
\begin{equation}
	W_{q\bar{q}g} = 4 W_1 W_2 - \frac{4}{3} W_3
\end{equation}
where $W_1$, $W_2$ and $W_3$ are the wilson loops given in Fig. \ref{loopqqg}.

\begin{figure}
\begin{picture}(350,100)(0,0)
\put(100,0){\includegraphics[width=0.5\textwidth]{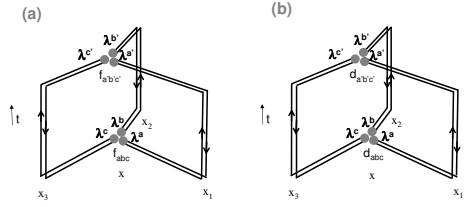}}
\end{picture}
\caption{Wilson loops for the three gluon system in the antisymmetric and symmetric color arrangements.}
\label{loopGGG}
\end{figure}

For the three gluons glueball the situation is more complex, since in this case there are two possible color wavefunctions.
One being antisymmetric for the exchange of two gluons and the other being symmetric. This comes from the fact that
the direct product of three adjoint representations of $SU(3)$ gives us two color singlets
\begin{equation}
	\mathbf{8} \otimes \mathbf{8} \otimes \mathbf{8} = \mathbf{1} \oplus \mathbf{1} \oplus ... 
\end{equation}
So, there are two different wilson loops for the three gluon glueball. One for the antisymmetric ($W_{3g}^A$) and the
other for the symmetric ($W_{3g}^S$) color arrangements. They are
\begin{equation}
	W_{3g}^A = f_{abc} f_{a'b'c'} \tilde{X}^{aa'} \tilde{Y}^{bb'} \tilde{Z}^{cc'}
\end{equation}
\begin{equation}
	W_{3g}^S = d_{abc} d_{a'b'c'} \tilde{X}^{aa'} \tilde{Y}^{bb'} \tilde{Z}^{cc'} ,
\end{equation}
where $\tilde{X}$, $\tilde{Y}$ and $\tilde{Z}$ are the three adjoint paths corresponding to three gluons creation,
propagation and annihilation shown in Fig. \ref{loopGGG}.

These two expressions could be simplified by using the relation
\begin{equation}
	\mbox{Tr}[ \lambda^a \lambda^b \lambda^c ] \lambda^a_{ij} \lambda^b_{kl} \lambda^c_{mn} =
	\frac{16}{9} \delta_{ij} \delta_{kl} \delta_{mn}
	- \frac{8}{3} \delta_{ij} \delta_{lm} \delta_{nk}
	- \frac{8}{3} \delta_{il} \delta_{jk} \delta_{mn}
	- \frac{8}{3} \delta_{in} \delta_{jm} \delta_{kl}
	+ 8 \delta_{il} \delta_{jm} \delta_{kn} ,
\end{equation}
which gives us
\begin{eqnarray}
\label{antisymmetricoperator}
	W_{3g}^{A} &=&  4 Tr[ X Y^\dagger ] Tr[ Y Z^\dagger ] Tr[ Z X^\dagger ] 
	+ 4 Tr[ X^\dagger Y ] Tr[ Y^\dagger Z ] Tr[ Z^\dagger X ]
	\\ \nonumber &&
		- 4 Tr[ X Z^\dagger Y X^\dagger Z Y^\dagger ] - 4 Tr[ X Y^\dagger Z X^\dagger Y Z^\dagger ]
\end{eqnarray}
and
\begin{eqnarray}
	W_{3g}^S &=& 4 Tr[ X Y^\dagger Z X^\dagger Y Z^\dagger ] + 4 Tr[ X^\dagger Z Y^\dagger X Z^\dagger Y ]
\nonumber \\ &&
	- \frac{16}{3} Tr[ X Y^\dagger ] Tr[ X^\dagger Y ]
	- \frac{16}{3} Tr[ Y Z^\dagger ] Tr[ Y^\dagger Z ]
	- \frac{16}{3} Tr[ Z X^\dagger ] Tr[ Z^\dagger X ]
\nonumber \\ &&
	+ 4 Tr[ X^\dagger Y ] Tr[ Y^\dagger Z ] Tr[ Z^\dagger X ] + 4 Tr[ Y^\dagger X ] Tr[ Z^\dagger Y ] Tr[ X^\dagger Z ]
	+ \frac{32}{3} \ .
\label{symmetricoperator}
\end{eqnarray}

\begin{figure}
\begin{picture}(350,100)(0,0)
\put(0,100){\includegraphics[width=0.5\textwidth,angle=-90]{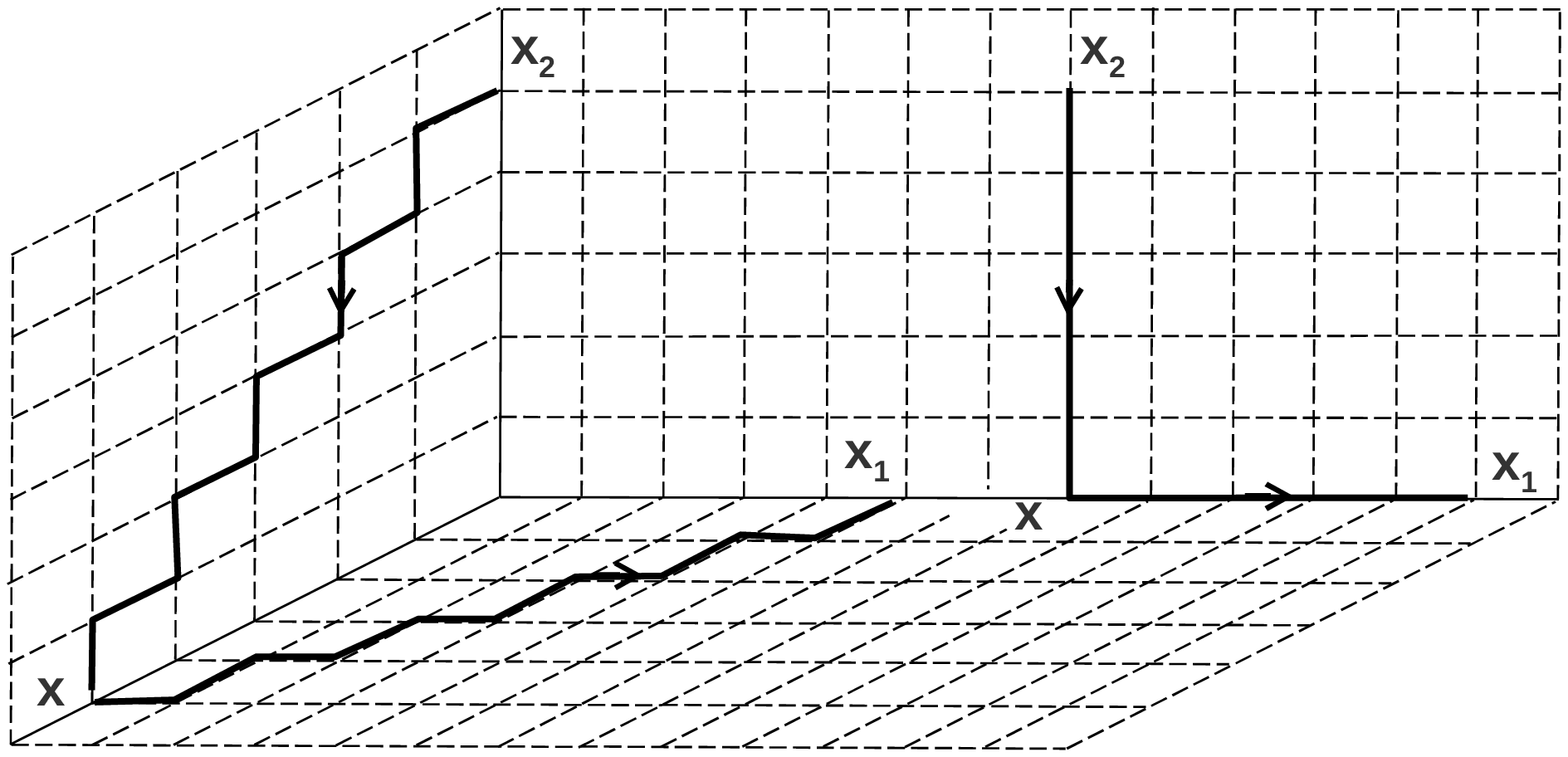}}
\put(220,100){\includegraphics[width=0.5\textwidth,angle=-90]{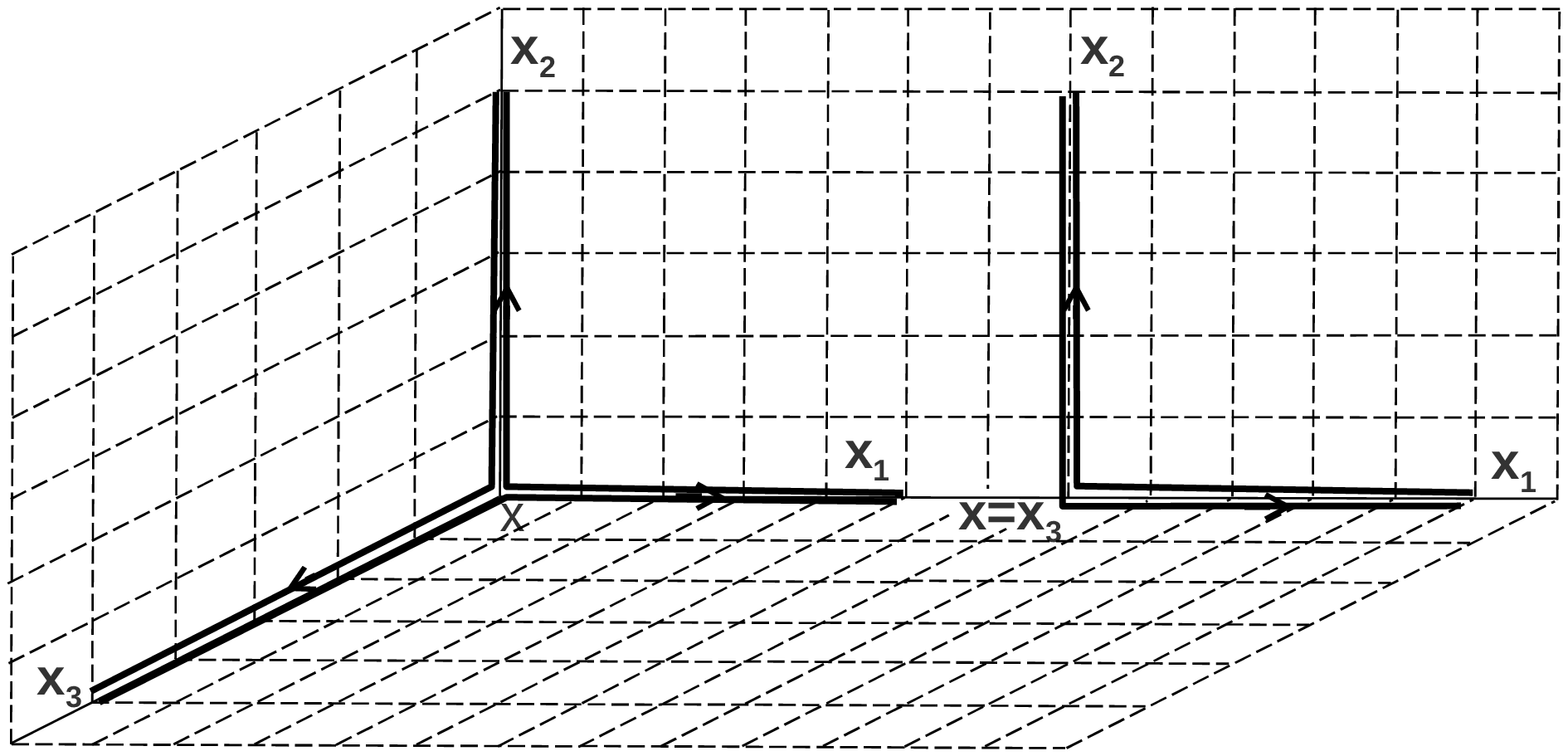}}
\end{picture}
\caption{Left: On axis and $45^o$ directions used in the computation of $W_{q\bar{q}g}$.
Right: Geometries used in the case of the three gluon glueball operator.}
\label{paths}
\end{figure}

In the computation of the $q\bar{q}g$ wilson loop we use, not only the on-axis directions, but also the $45^o$ off-axis
directions (see Fig. \ref{paths}). With this, we can, not only calculate the static potential for angles of
$0^o$, $90^o$ and $180^o$, but also angles of $45^o$, $60^o$, $120^o$ and $135^o$.

For the computations of the three-gluon wilson loops, we use two geometries (Fig. \ref{paths}):
One in which the three gluons form an equilateral triangle and, another, in which they form an rect isosceles
triangle. In the first one, the three gluons are in the positions $(r,0,0)$, $(0,r,0)$ and $(0,0,r)$ and join in $(0,0,0)$.
In the second, we have one of the gluons in the origin (coinciding with the meeting point) and the other two in the axis,
for instance in $(r,0,0)$ and $(0,r,0)$.

The static potential are extracted, as usual, by fitting the wilson loop mean value, at large euclidean times,
by a decaying exponential

\begin{equation}
	W(t) = C e^{ - V t }
\end{equation}

Our simulation uses the pure gauge $SU(3)$ Wilson action. For a lattice $24^3 \times 48$ and 
$\beta = 6.2$, 141 configurations  were generated, via a combination of Cabbibo-Mariani and overrelaxed updates, with the version 6 of the MILC code \cite{MILC}.

For the hybrid gluon we compute the hybrid potential as a function of the angle between the quark-gluon and the antiquark-gluon
segments $\theta$ and the lengths of the two segments ($r_1$ and $r_2$).

For the three gluon glueball the static potential is computed as a function of perimeter of the triangle with vertices in the
three gluons.

\section{Models of confinement}

\begin{figure}
\centering
\includegraphics[width=0.7\textwidth,angle=-90]{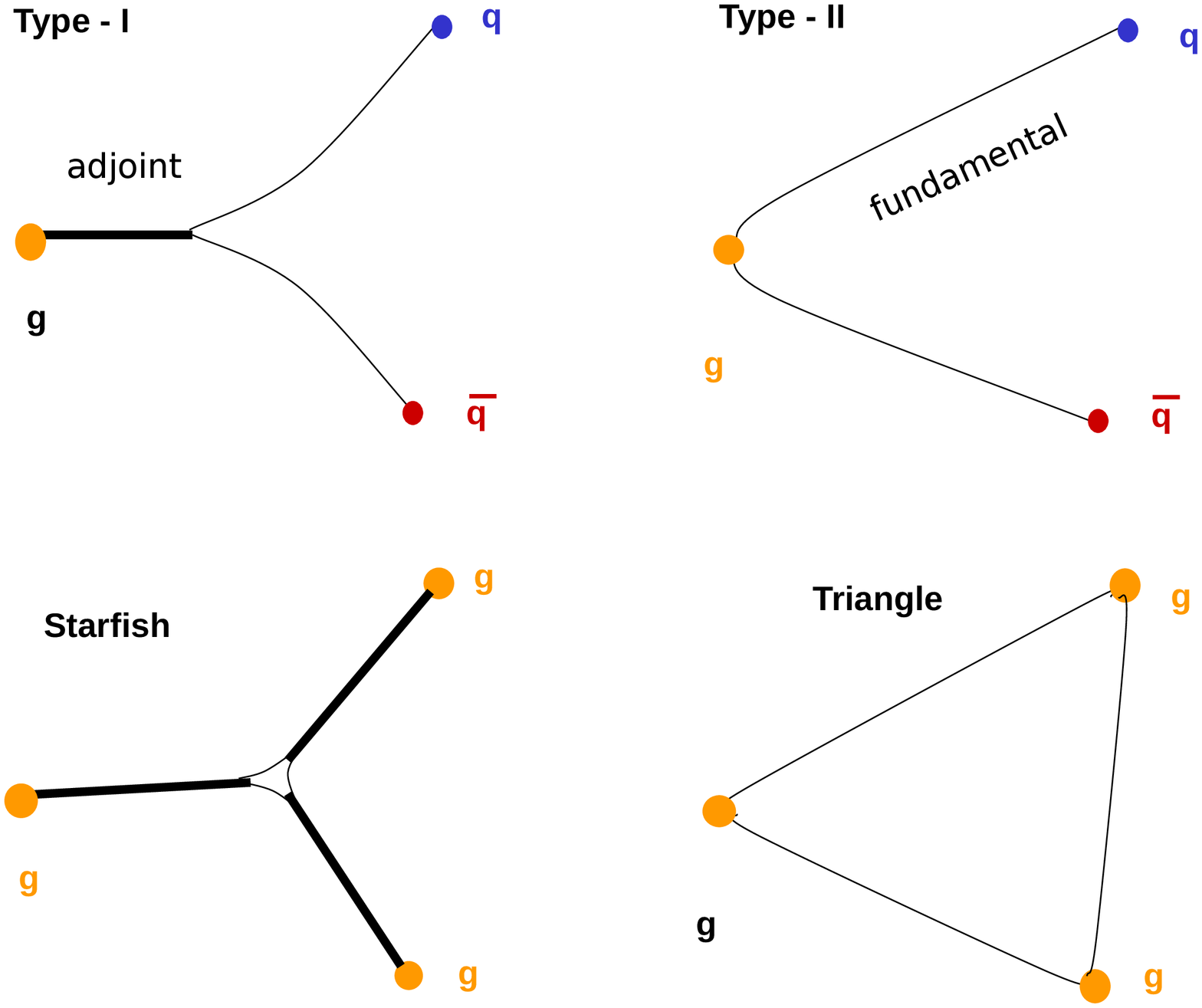}
\caption{The two models of the two models of confinement: type I and type II.}
\label{type1type2}
\end{figure}

We compare different models of confinement.
The simplest one is the Casimir scaling model, in which the static potential is given by a sum of two body potentials,
with each two-body potential being proportional to $\lambda_i \cdot \lambda_j$.
There are also models of confinement based on type I and type II superconductors. In the type I superconductor model, the fundamental strings tend to fuse giving origin to adjoint strings, such as as in a type I superconductor, single vortices fuse, giving origin to excited vortices. Similarly, there is a type II superconductor model, in which, the fundamental strings are kept apart and don't create excited strings, as in a type II superconductor, where two single vortices are more stable than an double vortex (see Fig. \ref{type1type2}).
For the case of the $q \bar{q} g$ the type-II model corresponds to having two independent strings for the
quark-gluon and antiquark-gluon links, and in the type-I, the two string fuse originating an adjoint string.
In the three-gluon system, the difference between the two geometries is more clear. In the case of the type-I model, each
gluon should link to the other two by straight fundamental strings, giving rise to a triangular geometry of the strings.
In the type I model, the strings form an adjoint string linking the three gluons, in a geometry similar to that of the
fundamental string in a baryon . 
This geometry we call starfish geometry (see Fig. \ref{type1type2}).

\section{Results for the hybrid meson}

\begin{figure}
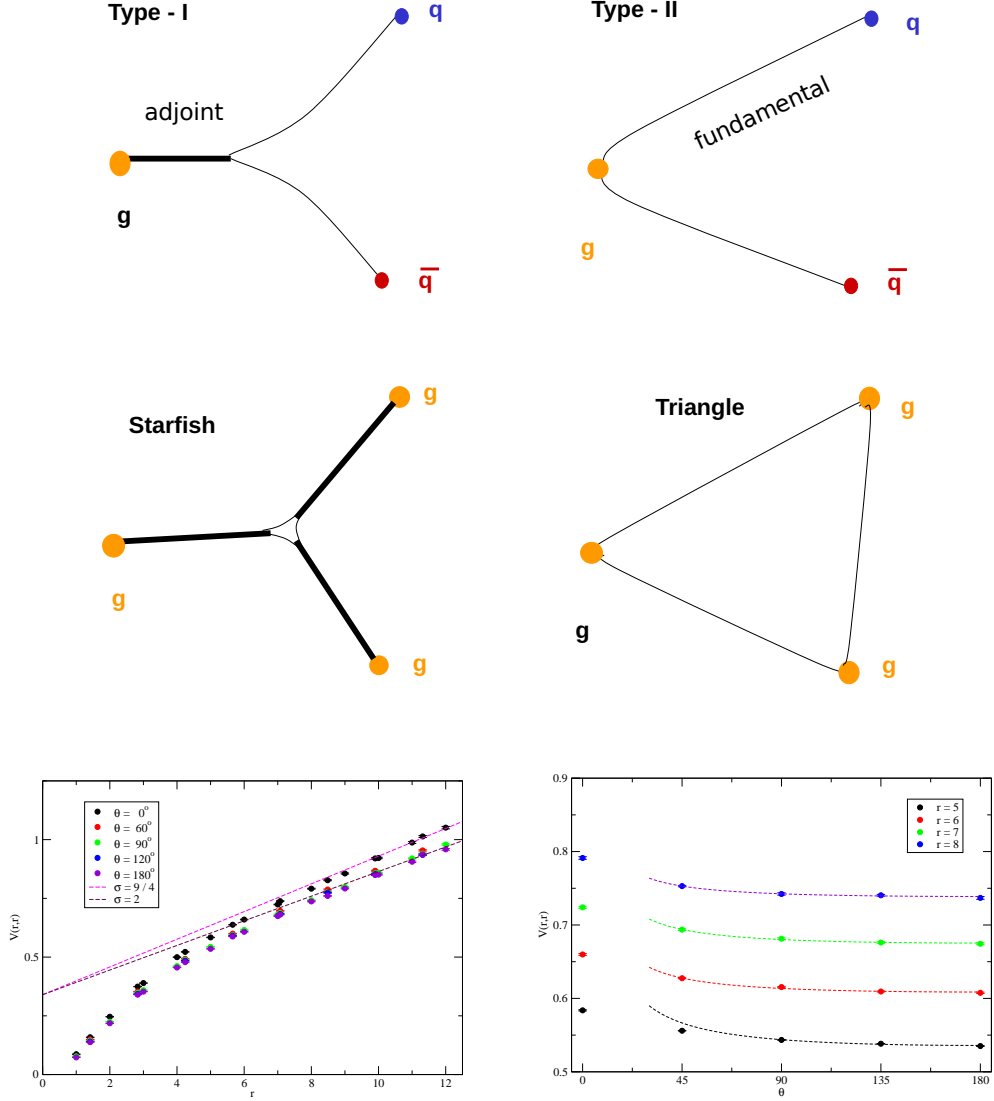

\begin{picture}(350,100)(0,0)
\put(20,0){
\includegraphics[width=0.4\textwidth]{VRRsigma.eps}
}
\put(220,0) {
\includegraphics[width=0.4\textwidth]{qqgcoul.eps}
}
\end{picture}
\caption{Left: Results for the V(r,r) on the hybrid meson for various angles and comparision with two different string tensions. Right: Results for V(r,r) as a function of $\theta$ for different values of $r$ }
\label{vrr}
\end{figure}

In Fig. \ref{vrr} we can see the results for the hybrid meson for the special case where $r_1 = r_2$ for different
value of the angle $\theta$ and there are also ploted two lines, with slopes $2 \sigma$ and $ \frac{9}{4} \sigma $.
In the case of angles $\theta \geq 60^o$ we see that the potential aproaches the $2 \sigma$ slope, meaning that, for those
angles we have, essentialy, two independent fundamental strings. However for $\theta = 0$, the graph approaches the 
$\frac{9}{4} \sigma$ slope, meaning that for this angle, the two fundamental strings collapse into one adjoint string.

We also see the results for the case $r_1 = r_2 = r$ as a function of $\theta$ for different values of
$r$. This results are fitted to a coulomb like potential, which is the only potential that changes with $\theta$ in the
model of two independent strings (type II superconductor). As can be seen the Coulomb fits well the results for large
$\theta$ and $r$.

\section{Results for the three-gluon glueball}

\begin{figure}
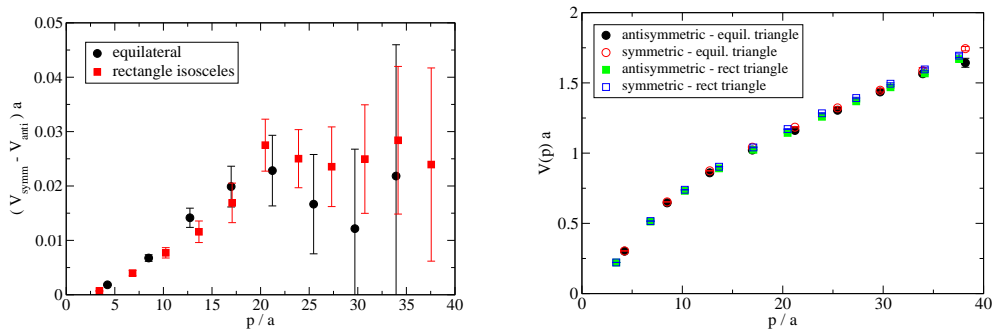

\begin{picture}(350,120)(0,0)
\put(20,0){
\includegraphics[width=0.4\textwidth]{Vdiff5.eps}
}
\put(220,0) {
\includegraphics[width=0.4\textwidth]{pot3g.eps}
}
\end{picture}
\caption{Left: Results for the difference of the three gluon static potential in the two color arrangements
 Right: Results for the three gluon potential as a function of the perimeter of the triangle formed by the three quarks. }
\label{pot3g}
\end{figure}

In the case of the three-gluon glueball we studied the difference between the potentials for the two color arrangements
$V_{symm} - V_{anti}$ and the potentials separately $V_{anti}$ and $V_{symm}$. Both are given in Fig. \ref{pot3g}.

The results for the difference $ V_{symm} - V_{anti} $ show that there is a systematic difference between the two potentials,
with the $V_{symm}$ being larger than $V_{anti}$. Fitting the difference between the two potentials to the form
$ V_{symm} - V_{anti} = C_0 + \sigma_{diff} r $, we see that we have $\sigma_{diff} = 0.04 \sigma$.

Fitting $V_{anti}$ and $V_{symm}$ to a potential of the form
$ V = C_0 - \alpha \sum_{i<j} \frac{1}{| \mathbf{r}_j - \mathbf{r}_i|} + \sigma' p $, where p is the perimeter gives us
$\sigma' = \sigma$. This show us that the geometry of the strings is the triangle geometry and not the starfish geometry, in
which case we would have $\sigma' = \frac{9}{4\sqrt{3}} \sigma$ for the equilateral triangle geometry, and
$\sigma' = \frac{9(1+\sqrt{3})}{8(1+\sqrt{2})}$ for the rect isosceles triangle geometry.

\section{Conclusions}

From the results for the hybrid meson and the three gluon glueball we conclude that confinement is essentially
realized by fundamental strings, as in a type II superconductor. 
This result, although simple, could be important for constituent quark and gluon models.
However, there are deviations from this behavior. Namely, in the hybrid meson, there is a repulsion between the two
fundamental strings (which link the quark to the gluon and the antiquark to the gluon), since when superposed they form
an adjoint with an energy superior to the energy they have separately. In the three gluons system, this also happens, since
when two of the gluons are on the same position, we get the case of the two gluon glueball, where an adjoint string links the
two gluons \cite{Bali:2000un}. So in both cases there is the formation of adjoint strings when there is a superposition of the fundamental strings. In the three gluon system there is, also, a phenomenon which the type II superconductor picture apparently does not account, which is the systematic difference between static potential of the two color arrangements.

\acknowledgments

Part of the present work was funded by the FCT grants 
PDCT/FP/63923/2005 and POCI/FP/81933/2007.

\end{document}